# A SIMPLE COSMOLOGICAL MODEL WITH DECREASING LIGHT SPEED

Juan Casado

Matgas Research Centre

Universidad Autonoma de Barcelona

Pl. Cívica, P12, 08193 Bellaterra (Barcelona) Spain

An alternative model describing the dynamics of a flat Universe without cosmological constant and allowing a gradual change of c with time is proposed. New relationships of redshift vs. distance and cosmic background radiation temperature are given. Values for the Universal radius, matter density, Hubble parameter, light deceleration, cosmic age and recombination time are obtained. Distant SNeIa faintness is explained within this decelerating, matter-dominated Universe without invoking dark energy. Horizon, flatness and other problems of standard Big Bang cosmology are solved without the need of inflation. The top speed of any signal, force, particle or wave at any time is limited by the expansion speed of the Universe itself.

## 1. Introduction

In recent years the possibility of non-standard cosmologies with varying light speed has been explored in order to solve cosmological issues such as the horizon, flatness, large scale homogeneity, initial fine-tuning or cosmological constant problems, among others (Ellis, 2000) as an alternative to the inflationary model. These theories have been recently reviewed (Magueijo, 2003) and its compatibility with the second law of thermodynamics has been studied (Youm, 2002). First attempts were from Moffat (1993). Albrecht and Magueijo (1999) postulated unchanged Friedmann equations and got a term proportional to $c'/c$, where $c'$ denotes the time derivative of c, the speed of light in vacuum. In those theories c undergoes a sudden change, from speeds many orders of magnitude larger than the present value ($c_o$) in a phase transition of the very early Universe, so that the initial value problems are avoided. Barrow (1999) extended these ideas to scenarios in which both c and G were proportional to some power of the scale factor. Clayton and Moffat (1998) implemented another varying speed of light model by considering a bimetric theory of gravitation in which one metric describes the standard gravitational vacuum and a second one describes the geometry in which matter is propagating. Moffat (2001, 2002, 2003) has discussed some cosmological implications of his different models. Usually in these works the period of variation of c is confined to the early Universe, the cosmological constant remains in the equations, some new parameters with non specified values are used and fitting of each theory with astrophysical observations is not discussed.

On the other hand, it has been theoretically shown how the speed of low-energy photons could be higher than c, depending on the energy density of modified vacua (Latorre et al. 1995). Kiritsis (1999) concluded that when a test brane moves in a black hole bulk space-time, c varies as the distance from the brane to the black hole. Alexander (2000) generalised this model by including rotation and expansion of the bulk so that c gets stabilised at long times.

In this paper a concrete model with smoothly decreasing light speed is presented. This model yields exact solutions for cosmological dynamics and avoids the use of free parameters or free functions to fit the observations. The model needs no phase transitions nor cosmological constant. In section 2 the fundamental assumptions and postulates are developed into some simple equations. In section 3 the redshift-distance relationship is revised and quantitative results from our model are compared with observational data. In section 4 the present light deceleration value is obtained. In section 5 Jordan's adimensional numbers are discussed in light of the present model. Section 6 shows that the flatness and the horizon problems can be avoided without inflation. Finally, in section 7 a new relationship between cosmic background radiation temperature and redshift, which can be used to test the present model, is given.

## 2.Fundamentals

Through this work, the specific time t has been chosen to be the comoving proper time. Speed of light depends on time, but not on the position or the velocity of the observer. The Universe is assumed to be spatially homogeneous and isotropic and practical flatness of space is also assumed.

It is remarkable that the farthest observable events, including high redshift galaxies and photons from cosmic background radiation (CBR), originated at distances close to ca. 15 billion light-years, the estimated Hubble radius of the Universe today ($R_o$). In other words, we can reach practically the entire history of the Universe by observing very distant events (except for the first ca. 300,000 years when it was opaque). This could seem obvious, but consider for a moment an Universe expanding 10 times more slowly than ours: then the most distant objects would be 'only' about 1.5 billion light-years away. On the other hand, if the Universe had expanded 10 times faster than it does, then most quasars, distant galaxies and CBR would be out of our event horizon. Therefore it seems that the maximum expansion speed, i.e. the rate at which the Hubble radius varies in time as reflected by the present Hubble parameter ($H_o$), is similar to the speed of light today ($c_o$):

$$H_o = \frac{R'_o}{R_o} \approx \frac{c_o}{R_o}, \qquad (1)$$

where $R'_o$ is the time derivative of $R_o$. Then $R'_o$, the actual maximum speed of universal expansion, is approximately $c_o$. Let's look at this point in a different way. Consider first the possibility that $c_o > R'_o$, then photons could escape from the Universe, which of course does not happen. Another possibility would be $c_o < R'_o$, but then two objects very far away in the Universe could recede from each other faster than $c_o$, which has never been observed and would violate relativity. Thus, both possibilities have no physical sense because they lead to absurdities. Furthermore, if the entire Universe is causally interconnected then its maximum expansion rate nowadays must be $c_o$ (see discussion on the horizon problem below). Therefore we assume $c_o = R'_o$, i. e. this expansion speed practically matches the speed at which the event horizon recedes from us. In more general terms, for different moments of the Universe history we have:

$$R' = c. \qquad (2)$$

This generalisation is justified because the above arguments should be valid in any cosmic time and, on the other hand, it would be much too coincidence if equation 1 only holds for the present Universe. An equivalent equation was already noticed by Pascual Jordan decades ago as will be discussed in section 5.

From (2) it is immediate that:

$$\frac{R'}{R} = H = \frac{c}{R} \qquad (3)$$

From the field equations of general relativity (GR), the critical density of the Universe reads:

$$\rho_c = \frac{3H^2}{8\pi G} = \frac{3c^2}{8\pi GR^2}.$$  (4)

On the other hand, an Universe of total gravitational mass M with spherical symmetry would have a density:

$$\rho = \frac{3M}{4\pi R^3}$$  (5)

Then, for an Universe with critical density we obtain from (4) and (5):

$$1 = \frac{Rc^2}{2GM},$$  (6)

which agrees with the classical Schwarzschild solution for GR describing a spherical black hole (equation (6) for black holes was also derived by Laplace from classical gravitation). So, the empirical observation of a universal density of matter close to the predicted for a black hole of radius $R_o$ (Singh, 1974; Sidharth, 2000; Casado, 2002) would not be a mere coincidence for present-day Universe, but a prediction of this model for practically any cosmic time. Therefore, in a flat Universe that follows (2), if R grows, then c decreases; i. e. as this Universe expands, c should decrease, so that both terms in equation

$$2GM = Rc^2$$  (7)

will remain constant (assuming that the gravitational constant G and the universal mass M **are real constants**). In other words: Could a static equation like (7) hold for an expanding Universe? The answer is yes only if c is continuously decreasing.

In spite of the existing literature on varying speed of light theories, this heterodox postulate should be further justified. Einstein himself disclaimed a unlimited validity of c as an universal constant, even if this was a postulate of special relativity. In his own words, before the first publication of GR (Einstein, 1912):

"the constancy of the velocity of light can be maintained only insofar as one restricts oneself to spatio-temporal regions of constant gravitational potential..."

And afterwards (Einstein, 1920):

"...according to the general theory of relativity, the law of the constancy of the velocity of light in vacuo, which constitutes one of the two assumptions in the special theory of relativity and to which we have already frequently referred, cannot claim any unlimited validity. A curvature of rays of light can only take place when the velocity of propagation of light varies with position."

It also has to be stressed that the present model includes the 'weak equivalence principle' stating that the trajectory of any freely falling body or particle does not depend on its internal structure, mass or composition. Furthermore, our model also respects the local Lorentz invariance. It only disagrees with the stronger or 'Einstein equivalence principle' in that any non-gravitational experiment may not be independent of **when** it is performed.

At which rate would c decrease? To solve this question we have to introduce an equation relating R and t. The simplest one derives from classical gravitation, which is a good approximation if our Universe is practically flat, as all the observations seem to indicate, and finite. In a Universe being spatially flat ($k=0$, $\Omega=1$), matter-dominated ($p\approx 0$, see section 7) and without cosmological constant ($\Lambda=0$), the Friedmann equations agree with classical gravitation. This was also the case in the Einstein-De Sitter cosmology.

So, by Newton laws divided per mass of test particle:

$$R'' = \frac{-GM}{R^2}, \qquad (8)$$

where R'' denotes the second derivative of R vs. t, i.e. the gravitational deceleration of Universal expansion.

Now, combining (2) and (8):

$$R'' = \frac{dc}{dt} = \frac{-GM}{R^2}. \tag{9}$$

Substituting R by its expression from (7) we have:

$$\frac{dc}{dt} = \frac{-GMc^4}{4G^2M^2} = \frac{-c^4}{4GM} \tag{10}$$

$$-\frac{dc}{c^4} = \frac{dt}{4GM}, \tag{11}$$

and integrating we obtain:

$$\frac{1}{3c^3} = \frac{t}{4GM}, \tag{12}$$

so that:

$$c^3 = \frac{4GM}{3t} \tag{13}$$

Then, substituting back R instead of its expression from (7), we get:

$$c = \frac{4GM}{3c^2t} = \frac{2R}{3t}, \tag{14}$$

and thus:

$$R = \frac{3ct}{2}. \tag{15}$$

Now it is immediate to deduce

$$R^3 = \frac{9\,GMt^2}{2} \tag{16}$$

These simple expressions relate R, c and t, but we still cannot obtain c', unless we determine R or t independently. We address this issue by derivation from empirical values of redshift (z) and distances in the next section.

### 3. Redshift-distance relationship

Modern astrophysics attributes the cosmological redshift to the stretching of wavelengths of photons as they propagate in an expanding Universe:

$$z = (\lambda_o/\lambda)-1 = (R_o/R)-1. \tag{17}$$

So that z values depend essentially on the scale factor $R_o/R$ (i.e. the ratio of Universal radii when the light was received and emitted respectively) and do not imply any specific velocity of expansion, opposite to classical Doppler interpretation. That opens the possibility of having a parameter $H_o$ different of nowadays accepted values.

Within our model the cosmological redshift is basically gravitational and is observed as a decrease of photons frequency as light travels against the Universal gravitational field when going from the past, more dense Universe, to the present one, just as any other particle losses energy during the expansion. It has to be noticed that only the observable Universe has effects, such as forces, on us, and the observable Universe belongs to the past, so that we are feeling gravitational forces coming from the past in the same way we are receiving light from past events, given that speed of gravity propagation is also c (see section 6). This redshift is also defined as usually:

$$z = (\nu - \nu_o)/\nu, \qquad (18)$$

where $\nu = c/\lambda$. Since $\lambda \propto R$ and, according to (7), $c \propto R^{-1/2}$, $\nu$ should be proportional to $R^{-3/2}$. Then:

$$z = \frac{R^{-3/2} - R_o^{-3/2}}{R_o^{-3/2}} = \left(\frac{R_o}{R}\right)^{3/2} - 1, \qquad (19)$$

a relationship different to conventional Eq. (17). Making $R_o - R \approx r$, the observed distance to a certain object, and rearranging we can linealize (19) to:

$$f(z) = 1 - (z+1)^{-2/3} \approx r/R_o \qquad (20)$$

When f(z) is plotted vs. distance (Tully, 1988) one obtains a good straight line ($R^2 = 0.996$) up to distances over 7 billion of (conventional) light-years with intersection of the axis close to the origin (figure 1).

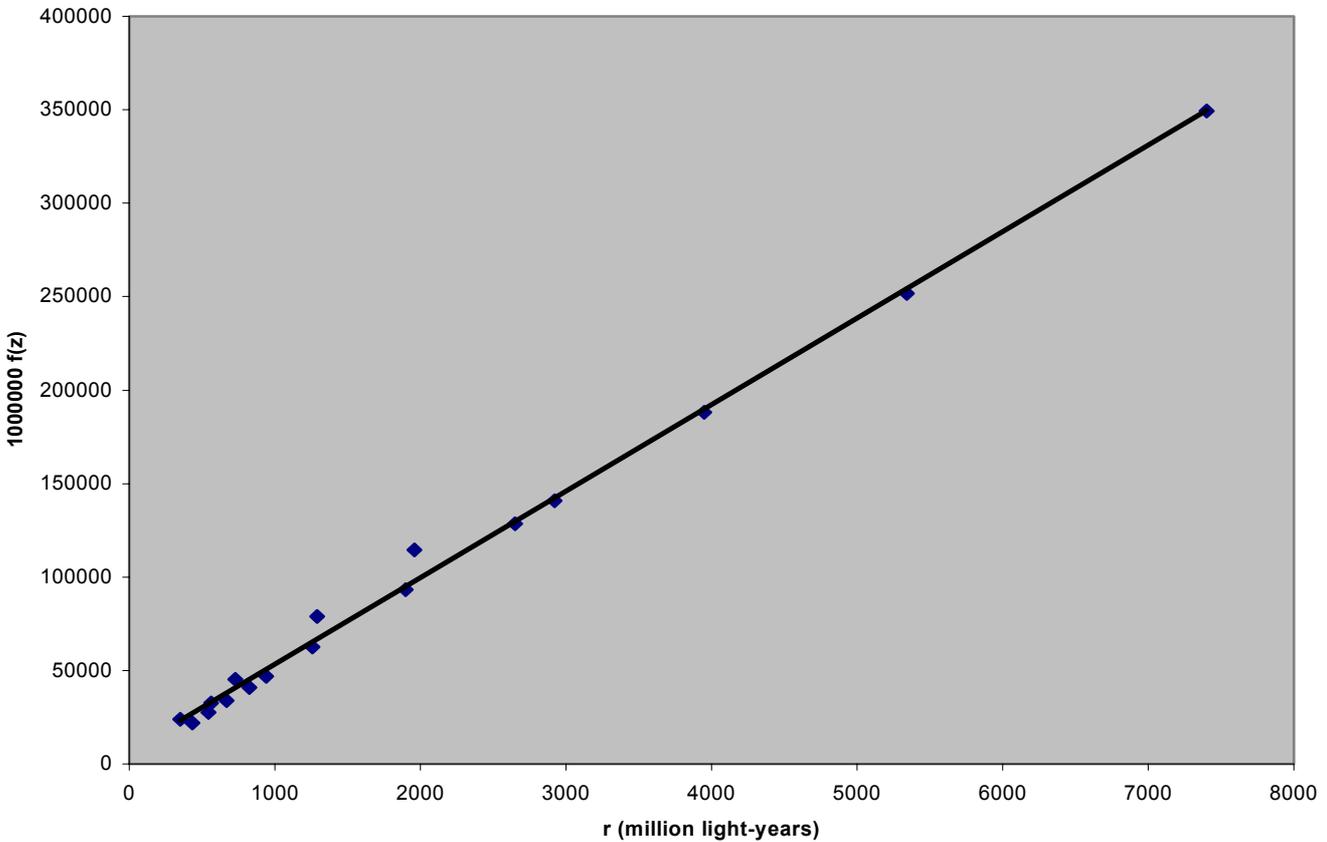

*Figure 1*. The function f(z) of equation (20) vs. distance r for 17 galaxies and quasars. The Y scale has been conveniently expanded by a factor $10^6$.

The slope of that line is $1/R_o$ and the resulting value for $R_o$ is close to $2.0 \ 10^{26}$m, a factor 3/2 bigger than the value predicted by the standard theory (Weinberg, 1972). Our model predicts -without need of introducing an 'ad hoc' repulsive dark energy to accelerate Hubble expansion- that supernovae with z ~ 1 are farther away than previously expected, so that they will appear fainter, as recently observed (Riess et al., 1998, 2001). Within our model an object showing z = 1 would be observed at a distance of $7.4 \ 10^{25}$ m, while within the standard model it would be at $6.6 \ 10^{25}$ m. This difference actually leads to about 25% less apparent luminosity than previously expected. Our hypothesis can be tested through further SNeIa data at different redshifts. If true, the cosmic expansion might not be accelerating after all.

Now we can apply equation (15) to obtain, for the present Universe:

$$t_o = \frac{2 R_o}{3 c_o} = 4.5 \ 10^{17} s = 1.4 \ 10^{10} \text{ years}, \qquad (21)$$

very close to the accepted value, but obtained in a different way.

With these quantitative results we can easily obtain the present matter density by calculating the Hubble parameter from:

$$H_o = \frac{c_o}{R_o} = \frac{2}{3 t_o} = 1.5 \ 10^{-18} \ s^{-1}, \qquad (22)$$

and substituting it on equation (4). The result is $\rho_m \approx 4 \ 10^{-30}$ g/cm$^3$. This value is different to the conventional critical density, but agrees quantitatively with the matter density in standard model with $\Omega_m \approx 0.4$. According to our model, the present matter density equals the corrected critical density obtained from the value of the Hubble parameter in (22). There is no room for $\Omega_\Lambda$. In connection with this, let us recall that $H_o$ is a crucial cosmological parameter that, however, has been very poorly determined, with values that had varied historically as much as one order of magnitude. The value in Eq. (22) agrees with the lowest experimental determinations obtained, but is significantly lower than the currently accepted values (ca. 70 km s$^{-1}$ Mpc$^{-1}$ = 2.3 10$^{-18}$ s$^{-1}$)

Notice that equation H = 2/(3t) is in agreement with the Einstein-De Sitter model, but now the figures are different. The accepted values of $H_o$ would lead, within that model, to an Universe of less than 10 billion years, younger than the oldest stars and globular clusters (with more than 12 billion years). This oldness problem contributed to discard the Einstein-De Sitter cosmology. The present model solves this problem given the lower values of $H_o$. Equation (22) can also be generalised to other moments in the cosmic history.

### 4. Present value of light deceleration

Finally, from the radius, the density and equation (9) is now easy to obtain the present value of light deceleration, which results to be 2.2 10$^{-10}$ m/s$^2$ , i. e. our model predicts that a decrease of 1m/s could be observed in about 140 years. This very small value has not been observed, but could be hidden within the error bars in recent determinations of c. Very precise laboratory measurements of c could detect this deceleration in a few decades.

In connection with this, it is remarkable that, using precision lunar orbital periods from 1978 to 1981, Van Flandern (1984) obtained a small deceleration in c:

$$-c'/c = (3.2 \pm 1.1)\ 10^{-11}/\text{year}. \quad (23)$$

This result represents a light deceleration of about 3 10$^{-10}$ m/s$^2$ , in agreement, taking into account the error margins quoted, with our calculated value.

The variation of c should be tested with 'mechanical' clocks such as those based on mechanical vibrators, pulsars rotation or planetary revolution, because atomic clocks periods depend on c.

Possibly this model could be also developed in terms of time dilatation instead of c decrease, but then the equations involved would be not so simple as the next section shows.

## 5. Jordan's adimensional numbers

Let us now recall an old numeric approach to cosmology. Probably inspired by Dirac, Jordan combined 6 important cosmological magnitudes in order to obtain adimensional numbers (Singh, 1974). The magnitudes were:

$c$, the speed of light
$f$ the gravitational constant in GR: $f = 8\pi G/c^2$
$t_o$ the age of the oldest celestial object (close to the Universe age)
$\rho_o$ the present matter density in the universe
$H_o$ the Hubble parameter
$R_o$ a length 'constant' from Hubble galactic counts

He showed that only 3 independent adimensional numbers can be constructed combining them:

$$t_o H_o \sim 1 \qquad (24)$$

$$\frac{R_o}{ct_o} \sim 1 \qquad (25)$$

$$f \rho_o c^2 t_o^2 \sim 1 \qquad (26)$$

and, from observations or estimations of the magnitude values, concluded that all this numbers where about one. This amazing result led him to the interpretation of $R_o$ as the curvature radius of a Riemannian closed space. From (24) and (25) Jordan also concluded that $R_o$ was expanding at the speed of light since the Big Bang. So we see that the basic equation (2) can be also obtained from completely different grounds. Taking into account equations (5) and (25) and disregarding small numeric factors, he transformed (26) into the new expression:

$$fM \sim R_o. \tag{27}$$

From (27) an astonishing result was obvious: if R expands with cosmic time either $f$ or M should increase. According to our model, the interpretation is that $f$ is growing along with R, because c decreases. But Jordan did not consider this possibility.

Taking into account the expression for $f$ and rearranging (27) one can also obtain:

$$\frac{GM}{c^2} \sim R \tag{28}$$

an equation equivalent to (7) for Schwarzschild radius, except for the factor 2.

Within our model the exact adimensional numbers would be:

$$t\,H = 2/3 \tag{29}$$

$$\frac{R}{ct} = \frac{3}{2} \tag{30}$$

$$\frac{GM}{Rc^2} = \frac{1}{2} \tag{31}$$

These adimensional numbers describing our Universe are only derived from integral calculus and/or geometric factors. On the other hand, they should be valid not only nowadays but along the history of the Universe and therefore they describe the cosmological evolution of H, R, c and $\rho$.

Another immediate derivation of the present model is that the total mechanical energy of the Universe should be zero at any time, assuming by convention a vanishing potential energy at R = ∞

### 6. The flatness and the horizon problems

The overall flatness of the Universe, except for small local deformations of space-time near massive objects, is one of the key postulates of this model, as described in section 2.

Regarding the horizon problem, it vanishes as soon as the speed of propagation of light,

gravity and whatever other interactions are as fast as Universal expansion at any moment, so that the entire Universe emerging from the Big Bang singularity can be causally interconnected all the time (figure 2).

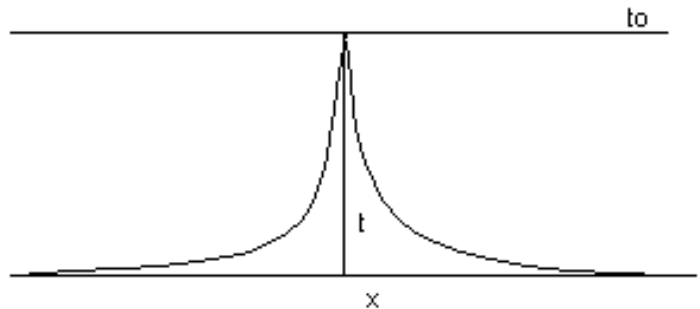

Fig. 2 Scheme of light hyperboloid showing the observable Universe; x is a normalised spatial co-ordinate and t is the temporal co-ordinate from the origin (t = 0), when the entire Universe was accessible, up to now (t = to). This is not a light cone anymore because c decelerates.

Inflation theory, not yet included in the standard model, also solves those problems (Guth and Lightman, 1997). It proposes that a very small fraction of second after the Big Bang, an exponential acceleration of the expansion took place so that the Universe reached superluminical expansion rates. Very shortly after, the expansion decreased again to a 'normal' Hubble flow. Superluminical expansion can be a problem within the theories of relativity because nothing should propagate faster than light. Inflationists argue that space-time frame itself could violate this principle but, in our physically observable world, space, time, matter and energy cannot be disentangled. For instance, it has no meaning to consider time without matter or energy since, in such a case, time could not be measured, i.e. time is nothing but a measure of the rate of changes in matter or energy. In a similar way, we can only measure spatial lengths of objects, distances among objects or distances travelled by energy waves; and we can do it only by using objects or waves. For related reasons it is impossible observing far away without observing time ago.

Furthermore, if one assumes that the shape of space-time can propagate faster than c, then gravity, described in GR as a curvature of space-time, could reach any point faster than light does, perhaps even instantaneously. But, on the other hand, it is commonly accepted that gravity propagates through waves that cannot be superluminical and, in fact, recent measurements of gravity speed agree with $c_o$ (Schewe et al., 2003). Therefore, if these measurements were confirmed, superluminical expansion of the Universe should be discarded.

Other shortcomings of inflation are the fine-tuning needed of the coupling constant in order to obtain the correct density profile in present Universe, the vacuum energy problem and the unnaturally flat potentials needed to solve the initial value problems (Moffat, 2002). Finally the low order quadrupole of the temperature anisotropy power spectrum from WMAP has lower amplitudes than expected from inflation (Efstathiou, 2003)

Within the present model the very fast expansion of the primitive Universe reveals to be not superluminical, although faster than present-day speed of light. Therefore the horizon problem is also solved without the need of an inflationary scenario. Neither cosmological constant nor dark energy, nor quintessence effects are required. In this way the cosmological constant/vacuum energy problem and the surprising coincidence of matter and dark energy dominance in present-day Universe (Caldwell et al., 1998) are also avoided.

From our perspective it is naturally understood why c is the maximum propagation speed for so many different phenomena such as waves, forces, particles, and any signal, interaction or causal connection between any two cosmic events. The reason appears to be surprisingly simple: since all of them happen within our Universe, they are limited to the maximum universal expansion rate: c.

## 7. The CBR temperature evolution

The present model agrees with the standard one in the way that temperature of the CBR has been decreasing as the scale factor increases, i.e.:

$$\frac{T}{T_o} = \frac{R_o}{R}, \qquad (32)$$

where $T_o$ denotes today temperature. The interpretation is however slightly different: T, which is a statistic measure of the kinetic energy of photons, is inversely proportional to R and, therefore, directly proportional to $c^2$, according to Eq. (6). This relationship seems reasonable since the kinetic energy of any particle is proportional to its squared velocity. In other words: the momentum *p* of any particle is proportional to its velocity, so that the momentum of a photon should be proportional to c and its energy (E = *p*c) will be proportional to $c^2$.

It is commonly accepted that recombination of nuclei and electrons to form atoms was allowed when the Universe temperature dropped to $T_c \approx 3000K$ (Weinberg, 1972), rendering the Universe transparent to electromagnetic waves. From that temperature we have that

$$\frac{T_c}{T_o} = \frac{3000}{2.73} = \frac{R_o}{R_c} \qquad (33)$$

So that this ratio is ca. 1100. On the other hand we have, taking into account Eq. (16):

$$\frac{t_o}{t_c} = \left(\frac{R_o}{R_c}\right)^{3/2} = 1100^{3/2} \qquad (34)$$

From this equation it is immediate that recombination time $t_c$ was $3.8 \cdot 10^5$ years after the Big Bang, in full agreement with NASA results obtained from WMAP by computational best fitting of parameters within standard model.

Combining equations (19) and (32) we obtain:

$$T = (z+1)^{3/2} T_o \qquad (35)$$

which gives the temperature of CBR in the past as function of z and $T_o$. This last equation differs from standard model (Ellis,2000). Both relationships can be tested by looking for excited molecules at high redshift. In fact this has been already done for a gas cloud at z = 2,34 (Srianand et al., 2000), finding a temperature range from 6.0 to 14K in agreement with standard model (predicting T=9.1K) but also with ours (T=6.1K). Again, further observations are needed to rule out at least one of them.

In standard model the energy density of the Universe is proportional to $T^4$ and inversely proportional to $R^4$ since the energy of each photon is considered to be proportional to T ($\propto 1/R$). In the present model this is also true, but when passed to mass equivalent density units, in order to compare with matter density, by dividing by $c^2$ ($\propto 1/R$), it results to be inversely proportional to $R^3$ and follows the same law that matter density. This statement implies that, as long as the ratio of photons to nucleons has been constant, the matter density has been about 9000 times higher than the energy density and our Universe has been matter-dominated along his history. This ratio justifies *a posteriori* the absence of a radiation pressure term in the above-derived equations that govern universal dynamics.

Further implications of the present model on the abundance of light elements, on the CBR fluctuation spectrum or in quantum cosmology will be addressed in future publications.

ACKNOWLEGMENTS

INCLUDEPICTUREINCLUDEPICTURE